\documentclass[allclo, smallextended, square, comma, sort&compress, natbib]{svjour3}
\bibpunct{[}{]}{;}{n}{}{,} 

\usepackage{amsmath}
\usepackage{amssymb}
\usepackage{graphicx}
\usepackage{color}
\usepackage{xspace}

\usepackage[mathscr,scaled=1.15]{urwchancal}
\DeclareFontFamily{OT1}{pzc}{}
\DeclareFontShape{OT1}{pzc}{m}{it}%
{<-> s * [1.15] pzcmi7t}{}
\DeclareMathAlphabet{\mathpzc}{OT1}{pzc}{m}{it}

\definecolor{purple}{rgb}{0.5,0,0.5}
\definecolor{blue}{rgb}{0.0,0,0.9}
\definecolor{prdblue}{rgb}{0.133,0.118,0.498}

\journalname{Few-Body Systems}
\begin{document}

\title{$N^\ast$ Structure and Strong QCD
}


\author{Craig D. Roberts
}


\institute{Craig D.\ Roberts \at
            Argonne National Laboratory, Argonne, Illinois 60439, USA.\\
              \email{cdroberts@anl.gov}
}

\date{
25 January 2018
}

\maketitle

\begin{abstract}
In attempting to match QCD with Nature, it is necessary to confront the many complexities of strong, nonlinear dynamics in relativistic quantum field theory, \emph{e.g}. the loss of particle number conservation, the frame and scale dependence of the explanations and interpretations of observable processes, and the evolving character of the relevant degrees-of-freedom.  The peculiarities of QCD ensure that it is also the only known fundamental theory with the capacity to sustain massless elementary degrees-of-freedom, gluons and quarks; and yet gluons and quarks are predicted to acquire mass dynamically so that the only massless systems in QCD are its composite Nambu-Goldstone bosons.  All other everyday bound states possess nuclear-size masses, far in excess of anything that can directly be tied to the Higgs boson.  These observations highlight fundamental questions within the Standard Model: what is the source of the mass for the vast bulk of visible matter in the Universe, how is its appearance connected with confinement; how is this mass distributed within hadrons and does the distribution differ from one hadron to another?  This contribution sketches insights drawn using modern methods for the continuum bound-state problem in QCD, and how they have been informed by empirical information on the hadron spectrum and nucleon-to-resonance transition form factors.
\end{abstract}

\section{Introduction}
\label{intro}
The existence of our Universe depends critically on the following empirical facts:
the proton is massive, \emph{i.e}.\ the mass-scale for strong interactions is vastly different to that of electromagnetism;
and it is absolutely stable, despite being a composite object constituted from three valence quarks;
but, on the other hand, the pion is unnaturally light, possessing a lepton-like mass, even though it is constituted from the same degrees-of-freedom that produce the proton.
These qualities are fundamental \emph{emergent phenomena}, the understanding of which is crucial if science is to reveal our origins, and could potentially be central to learning how the picture of Nature may be completed by moving beyond the Standard Model.

The Lagrangian defining quantum chromodynamics (QCD) can be written in a single line, with two definitions:
\begin{subequations}
\label{LQCD}
\begin{align}
{\mathpzc L}_{QCD} & = \bar q(x) \left[ \gamma\cdot D + {\mathpzc M} \right] q(x)
    + \tfrac{1}{4} G_{\mu\nu}^a(x)  G_{\mu\nu}^a(x)\,,\\
D_{\mu}^x &= \partial_\mu^x - i g \tfrac{1}{2} \lambda^a  B_\mu^a(x) \,,\\
G_{\mu\nu}^a(x) & = \partial_\mu^x B_\nu^a(x) - \partial_\nu^x B_\mu^a(x) + g f^{abc} B_\mu^b(x) B_\nu^c(x)\,,
\end{align}
\end{subequations}
where $g$ is the gauge coupling, $\{\tfrac{1}{2}\lambda^a|a=1,\ldots,8\}$ are the generators of $SU(3)$-colour in the fundamental representation, and $q(x)$ and $\{B_\mu^a(x)\}$ are quark and gluon fields.  ${\mathpzc M}$ in Eq.\,\eqref{LQCD} is a diagonal matrix whose entries are the Higgs-generated current-quark mass parameters, which provide the only explicit mass-scales in QCD.  For light hadrons (neutron, proton, pion, etc.), it is sufficient to consider ${\mathpzc M} = {\rm diag}[m_u,m_d]$.

Empirically, the mass-scale for the spectrum of strongly interacting matter is characterised by the proton's mass, $m_p \approx 1\,$GeV.  However, as just noted, the only apparent scales in chromodynamics are the current-quark masses; and in connection with everyday matter, these masses are just $1/250^{\rm th}$ of the empirical scale for strong interactions, \emph{i.e}.\ more-than two orders-of-magnitude smaller.  Crucially, no amount of ``staring'' at ${\mathpzc L}_{QCD}$ can reveal the origin of $m_p$.  This is a stark contrast to quantum electrodynamics (QED), \emph{e.g}.\ the spectrum of the hydrogen atom is set by $m_e\alpha^2$, where $m_e$ is the electron mass and $\alpha$ is the fine-structure constant, each of which is a prominent feature of ${\mathpzc L}_{\rm QED}$.

If one removes the current-quark mass from classical chromodynamics, the theory is scale invariant and the dilation current is conserved: $\partial_\mu {\mathpzc D}_\mu = T_{\mu\mu} = 0$, where $T_{\mu\nu}$ is the theory's  energy-momentum tensor.  There is no dynamics in a scale invariant theory, only kinematics.  Consequently, bound-states are impossible and our Universe can't exist.  This catastrophe is avoided by quantum effects.   In quantising QCD, the process of regularising and renormalising (ultraviolet) divergences introduces a mass-scale, an effect known as ``dimensional transmutation'', \emph{viz}.\ mass-dimensionless quantities become dependent on a mass-scale, and this entails the appearance of the chiral-limit ``trace anomaly'' \cite{Adler:1976zt,Collins:1976yq,Nielsen:1977sy} :
\begin{equation}
\label{SIQCD}
T_{\mu\mu} = \beta(\alpha(\zeta))  \tfrac{1}{4} G^{a}_{\mu\nu}G^{a}_{\mu\nu} =: \Theta_0 \,,
\end{equation}
where $\beta(\alpha)$ is QCD's $\beta$-function, $\zeta$ is the renormalisation scale, and this expression assumes the chiral limit for all quarks.

It is also worth emphasising that classical chromodynamics is a non-Abelian local gauge theory.  Consequently, the concept of local gauge invariance persists.  However, without a mass-scale there is no confinement.  For example, three quarks can be prepared in a colour-singlet combination and colour rotations will keep the three-body system colour neutral; but the quarks involved need not have any proximity to one another.  Indeed, proximity is meaningless because all lengths are equivalent in a scale invariant theory.  Hence, the question of ``Whence mass?'' is equivalent to ``Whence a mass-scale?'', which is equivalent to ``Whence a confinement scale?''.  It follows that understanding the origin of mass in QCD is quite likely inseparable from the task of understanding confinement.

Simply knowing that a trace anomaly exists does not deliver much: it only indicates that there is a mass-scale.  The crucial issue is whether or not one can compute and/or understand the magnitude of that scale.  It can certainly be measured, for consider the expectation value of the energy-momentum tensor in the proton:
\begin{equation}
\label{EPTproton}
\langle p(P) | T_{\mu\nu} | p(P) \rangle = - P_\mu P_\nu\,.
\end{equation}
In the chiral limit, it follows that
\begin{equation}
\label{anomalyproton}
\langle p(P) | T_{\mu\mu} | p(P) \rangle  = - P^2 = m_p^2  = \langle p(P) |  \Theta_0 | p(P) \rangle\,;
\end{equation}
and this supports a parton-model view that the entirety of the proton mass is produced by gluons: the trace anomaly is measurably large; and being directly identified with the square of the gluon field-strength tensor, that property should logically owe to gluon self-interactions, which are also responsible for asymptotic freedom.

There is a flip-side to Eq.\,\eqref{anomalyproton}, \emph{viz}.\ replace the proton state by the pion:
\begin{equation}
\label{anomalypion}
\langle \pi(q) | T_{\mu\nu} | \pi(q) \rangle = - q_\mu q_\nu
\Rightarrow     \langle \pi(q) | T_{\mu\mu} | \pi(q) \rangle \stackrel{\rm chiral~limit} = 0\,,
\end{equation}
because the pion is a massless Nambu-Goldstone mode in the chiral limit.
Evidently, the natural nuclear-physics mass-scale, $m_p$, emerges simultaneously with the apparent preservation of scale invariance in related systems.  Moreover, the expectation value of $\Theta_0$ in the pion is always zero, irrespective of the size of $m_p$.
Does this mean that the scale anomaly vanishes trivially in the pion state, \emph{i.e}.\ each term in the expression of the operator vanishes when evaluated in the pion and thus gluons contribute nothing to the pion mass?  The answer is NO.  Instead, as explained in Sec.\,\ref{Seclightquarks}, following Eq.\,\eqref{sigmaterm}, Eq.\,\eqref{anomalypion} owes to cancellations between different operator-component contributions; and the cancellation is exact in the pion channel because of dynamical chiral symmetry breaking (DCSB).  This is the content of Nambu's share of the 2008 Nobel Prize in Physics \cite{Nambu:2011zz}.

The problems of understanding the emergence of mass within the Standard Model, and elucidating its empirical and theoretical consequences define two of the most important challenges in basic science.  They can be addressed by studying elastic and transition form factors, and distribution amplitudes and functions, each of which provides a scale-dependent probe of hadron structure.  For example, they can be used to map the transition from the high-energy domain, within which ``Feynman's partons'' are the appropriate degrees of freedom, to the strong-QCD (sQCD) domain, wherein complex, dressed quasi-particles emerge and are required in order to provide a natural explanation of hadron properties.  Theoretically, the full machinery of renormalisable, relativistic quantum field theory is necessary to explain this metamorphosis and expose its signatures in observables.

Crucially, ``scaling'' is not a characteristic that is definitive QCD.  Such behaviour is practically a kinematic feature of cross-sections at large probe-momentum scales.  Rather, logarithms, scaling violations in the approach to a conformal limit, are the ``smoking gun'' for entrance into a domain where factorisation and hard-scattering formulae can be validated.  In this doorway the qualities of sQCD are expressed, \emph{e.g}.\ in the factors which define the shape of distributions and the overall scale of cross-sections.   Of course, correct numerical results alone are insufficient.  An understanding of the emergence of mass, confinement, etc., will only be found in a reductive explanation of the nature of the strong-to-weak QCD transition.

\section{Gauge Sector Dynamics}
Notably, all renormalisable four-dimensional theories possess a trace anomaly.  Hence, QED and QCD are alike in this; but years of comparing systems bound by electromagnetism with those produced by the strong interaction indicate that the size of the trace anomaly in QED must be very much smaller than that in QCD.  This disparity may be understood by noting that, with reference to the generating functional for one-particle irreducible Schwinger functions,
\begin{equation}
\label{vacpol}
\int d^4 x \, G_{\mu\nu}G_{\mu\nu} \sim \int d^4 x d^4 y \, A_\mu(x) D^{-1}_{\mu\nu}(x-y) A_\nu(y)\,,
\end{equation}
where $D^{-1}_{\mu\nu}$ is the fully-dressed gauge-boson 2-point function (Euclidean propagator).   Eq.\,\eqref{vacpol} highlights that if any mass-scale is to become associated with the trace anomaly, then it will be exhibited in the gauge-boson vacuum polarisation.
Textbooks show that the photon vacuum polarisation does not possess an infrared mass-scale; namely, $\Pi_{\rm QED}(k^2=0)=0$, and serves merely to produce the very slow running of the QED coupling, \emph{i.e}.\ any dynamical violation of the conformal features of QED is very small and hence the trace anomaly is negligible.
In contrast, owing to gauge sector dynamics, a Schwinger mechanism is active in QCD \cite{Aguilar:2015bud}, so that \begin{equation}
\left.k^2\Pi_{\rm QCD}(k^2)\right|_{k^2=0} \approx \Lambda_{\rm QCD}^2
\end{equation}
and the QCD trace anomaly expresses a mass-scale that is, empirically, very significant.
This sort of connection between the trace anomaly and a gluon mass-scale was first shown in Ref.\,\cite{Cornwall:1981zr}.  The intervening years have revealed a great deal about the infrared behaviour of the running coupling, dressed-gluon propagator and dressed-gluon-quark vertex; and the current state of understanding can be traced from an array of sources \cite{Aguilar:2015bud,Cornwall:1981zr,Boucaud:2011ug,Binosi:2014aea,Binosi:2016wcx,Binosi:2016nme}.

The fact that the gluon propagator saturates at infrared momenta, \emph{i.e}.\
\begin{equation}
\Delta(k^2\simeq 0) = 1/m_0^2, \quad m_0 \approx 0.5\,{\rm GeV} \approx \tfrac{1}{2} m_p\,,
\end{equation}
 is signficant.  It entails that the long-range propagation of gluons is dramatically affected by their self-interactions. Importantly, $m_0$
is a renormalisation-group-invariant (RGI) gluon mass-scale \cite{Ayala:2012pb,Binosi:2014aea,Binosi:2016wcx,Binosi:2016nme,Binosi:2016xxu,Cyrol:2016tym, Gao:2017uox}; and it is now known that although gluons act as massless degrees-of-freedom on the perturbative domain, they possess a running mass, whose infrared value is characterised by $m_0$.

The emergence of a gluon mass-scale reveals a new physics frontier within the Standard Model.  Asymptotic freedom ensures that QCD's ultraviolet behaviour is controllable; and $m_0> 0 $ entails that QCD dynamically generates its own infrared cutoff, so that gluons with wavelength $\lambda \gtrsim 1/m_0 \approx 0.5\,$fm decouple from the strong interaction, hinting at active confinement\footnote{The potential between infinitely-heavy quarks measured in numerical simulations of quenched lQCD -- the so-called static potential \cite{Wilson:1974sk} -- is disconnected from the question of confinement in our Universe.  The latter follows because light-particle creation and annihilation effects are essentially nonperturbative in QCD, so it is impossible in principle to compute a quantum mechanical potential between two light quarks \cite{Bali:2005fu,Prkacin:2005dc,Chang:2009ae}.  It follows that the flux tube measured in numerical simulations of lQCD with static quarks has no relevance to confinement in the light-quark realm of QCD.  Moreover, there is zero knowledge of the strength or extension of a flux tube between a static-quark and any light-quark because it is impossible even to define such a flux tube; and it is doubly impossible to define a flux-tube between a light-quark source and light-quark sink.  Hence, since the vast bulk of visible matter is constituted from light valence quarks, with no involvement of even an accessible heavy quark, the flux tube picture is not the correct paradigm for confinement in hadron physics.} \cite{Munczek:1983dx,Stingl:1985hx,Krein:1990sf,Burden:1991gd,Hawes:1993ef,Maris:1994ux,%
Bhagwat:2002tx,Roberts:2007ji,Bashir:2009fv,Bashir:2013zha,Qin:2013ufa,Lowdon:2015fig,Lucha:2016vte,Binosi:2016xxu}.  The existence of a running gluon mass, large at infrared momenta, has an impact on all analyses of the continuum bound-state problem and could also be a harbinger of gluon saturation \cite{Accardi:2012qut}.

\begin{figure}[t]
\centerline{\includegraphics[width=0.66\textwidth]{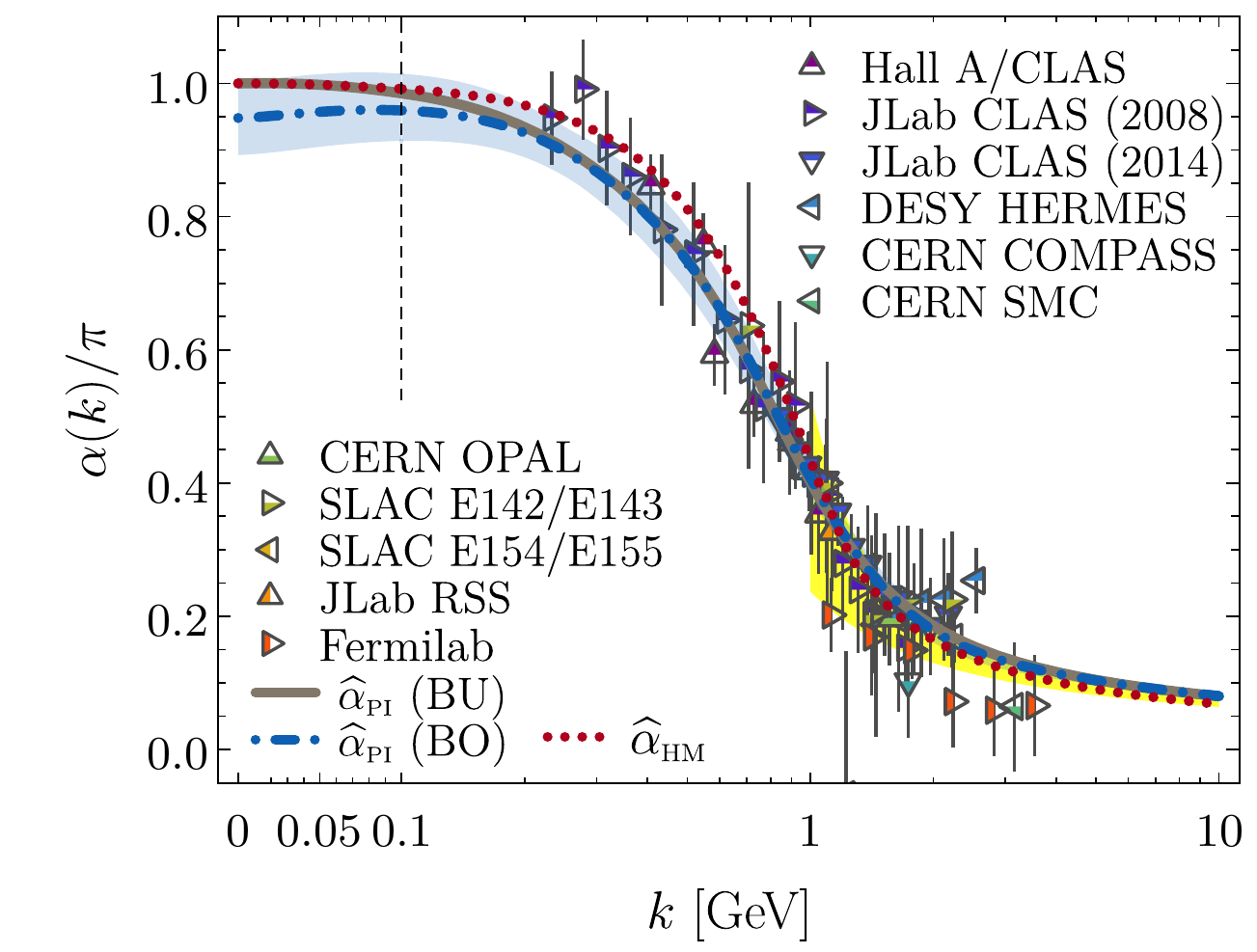}}
\caption{\label{FigwidehatalphaII}
Dot-dashed (blue) curve (BO): process-in\-de\-pen\-dent RGI running-coupling $\widehat{\alpha}_{\rm PI}(k^2)$ introduced in Ref.\,\cite{Binosi:2016nme}.  The shaded (blue) band bracketing this curve combines a 95\% confidence-level window based on existing lQCD results for the gluon two-point function with an error of 10\% in the continuum analysis of relevant ghost-gluon dynamics.
Solid (black) curve, (BU): update of the result in Ref.\,\cite{Binosi:2016nme}, including a previously-neglected ghost-loop contribution to the gluon self-energy, which evidently adds a little strength on $k\lesssim 0.2\,$GeV, as described in the Rodr{\'i}guez-Quintero \emph{et al}.\ contribution to these proceedings.
World data on the process-dependent effective coupling $\alpha_{g_1}$, defined via the Bjorken sum rule \cite{%
Deur:2005cf,Deur:2008rf,Deur:2014vea,%
Ackerstaff:1997ws,Ackerstaff:1998ja,Airapetian:1998wi,Airapetian:2002rw,Airapetian:2006vy,%
Kim:1998kia,%
Alexakhin:2006oza,Alekseev:2010hc,Adolph:2015saz,%
Anthony:1993uf,Abe:1994cp,Abe:1995mt,Abe:1995dc,Abe:1995rn,Anthony:1996mw,Abe:1997cx,Abe:1997qk,Abe:1997dp,%
Abe:1998wq,Anthony:1999py,Anthony:1999rm,Anthony:2000fn,Anthony:2002hy}.
The shaded [yellow] band on $k>1\,$GeV represents $\alpha_{g_1}$ obtained from the Bjorken sum by using QCD evolution \cite{Gribov:1972,Altarelli:1977,Dokshitzer:1977} to extrapolate high-$k^2$ data into the depicted region \cite{Deur:2005cf,Deur:2008rf}; and, for additional context, the dotted [red] curve is the effective charge obtained in a light-front holographic model, canvassed elsewhere \cite{Deur:2016tte}.
}
\end{figure}

There are many other consequences of the intricate nonperturbative nature of gauge-sector dynamics in QCD.  Important amongst them is the generation of a process-independent effective charge, $\widehat{\alpha}_{\rm PI}(k^2)$ -- see Ref.\,\cite{Binosi:2016nme} and the contributions of Brodsky, Deur and Rodr{\'i}guez-Quintero \emph{et al}.\ to these proceedings.  Depicted as the solid (black) curve in Fig.\,\ref{FigwidehatalphaII}, this is a new type of effective charge, which is an analogue of the Gell-Mann--Low effective coupling in QED \cite{GellMann:1954fq} because it is completely determined by the gauge-boson propagator.

The data in Fig.\,\ref{FigwidehatalphaII} represent empirical information on $\alpha_{g_1}$, a process-depen\-dent effective-charge \cite{Grunberg:1982fw} determined from the Bjorken sum rule, one of the most basic constraints on our knowledge of nucleon spin structure.  Sound theoretical reasons underpin the almost precise agreement between $\widehat{\alpha}_{\rm PI}$ and $\alpha_{g_1}$ \cite{Binosi:2016nme}, so that the Bjorken sum may be seen as a near direct means by which to gain empirical insight into QCD's Gell-Mann--Low effective charge.
Given the behavior of the prediction in Fig.\,\ref{FigwidehatalphaII}, it is evident that the coupling is everywhere finite in QCD, \emph{i.e}.\ there is no Landau pole, and this theory possesses an infrared-stable fixed point.  Evidently, QCD is infrared finite owing to the dynamical generation of a gluon mass scale.\footnote{%
A theory is said to possess a Landau pole at $k^2_{\rm L}$ if the effective charge diverges at that point.  In QCD perturbation theory, such a Landau pole exists at $k^2_L=\Lambda_{\rm QCD}^2$.  Were such a pole to persist in a complete treatment of QCD, it would signal an infrared failure of the theory.  On the other hand, the absence of a Landau pole in QCD supports a view that QCD is unique amongst four-dimensional quantum field theories in being defined and internally consistent at all energy scales.  This might have implications for attempts to develop an understanding of physics beyond the Standard Model based upon non-Abelian gauge theories \cite{Binosi:2016xxu,Appelquist:1996dq,Sannino:2009za,Appelquist:2009ka,Hayakawa:2010yn,%
Cheng:2013eu,Aoki:2013xza,DeGrand:2015zxa}.}

As a unique process-independent effective charge, $\widehat{\alpha}_{\rm PI}$ appears in every one of QCD's dynamical equations of motion, including the gap equation, setting the strength of all interactions.  $\widehat{\alpha}_{\rm PI}$ therefore plays a crucial role in determining the fate of chiral symmetry, \emph{i.e}.\ the dynamical origin of light-quark masses in the Standard Model even in the absence of a Higgs coupling.

\section{Light Quarks}
\label{Seclightquarks}
\begin{figure}[t]
\centerline{\includegraphics[clip,width=0.60\textwidth]{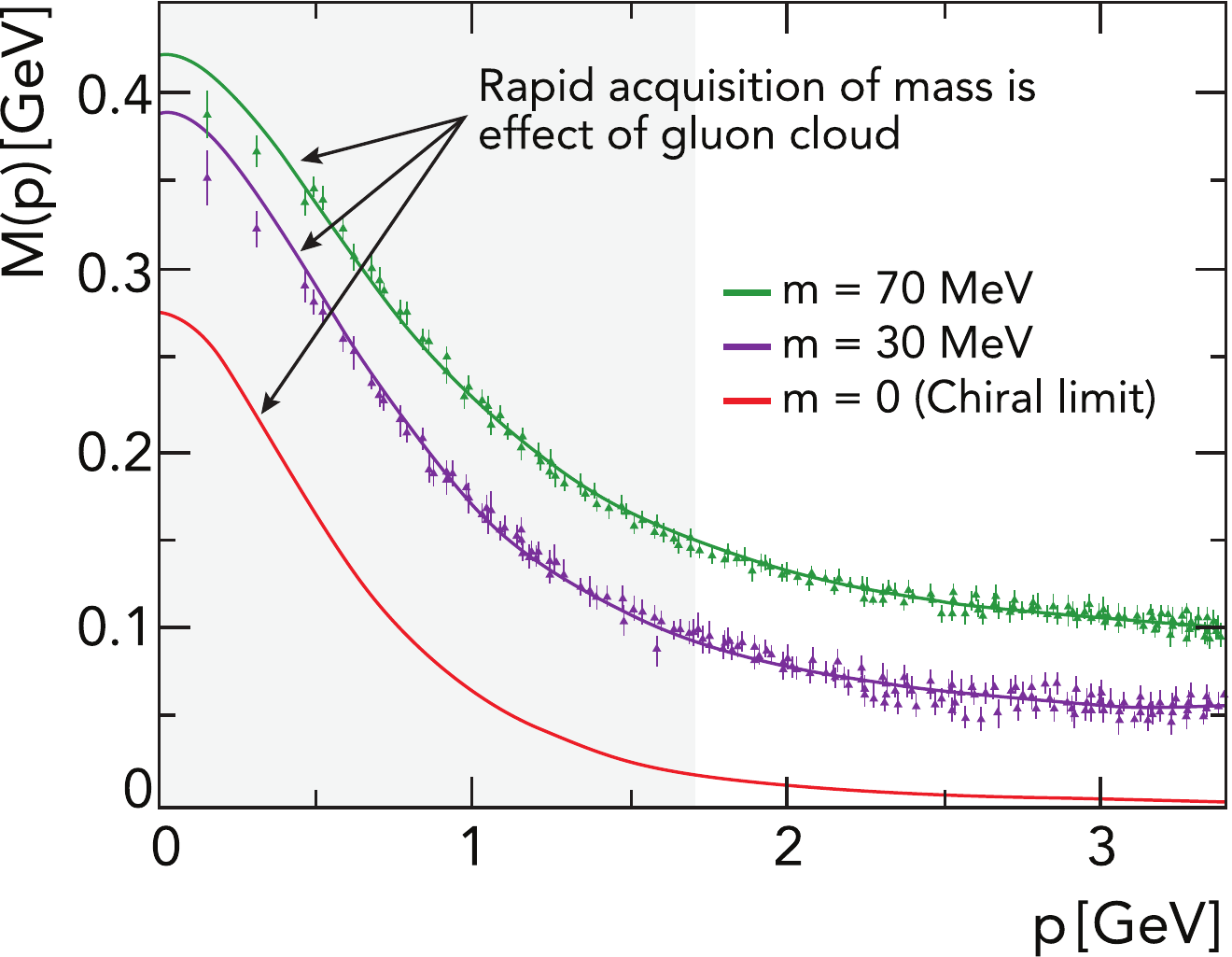}}
\caption{\label{gluoncloud}
Renormalisation-group-invariant dressed-quark mass function, $M(p)$ in Eq.\,\eqref{Spgen}: \emph{solid curves} -- gap equation results \cite{Bhagwat:2003vw,Bhagwat:2006tu}, ``data'' -- numerical simulations of lQCD \protect\cite{Bowman:2005vx}.
The current-quark of perturbative QCD evolves into a constituent-quark as its momentum becomes smaller.  The constituent-quark mass arises from a cloud of low-momentum gluons attaching themselves to the current-quark.  This is DCSB, the essentially nonperturbative effect that generates a quark \emph{mass} \emph{from nothing}; namely, it occurs even in the chiral limit.
Notably, the size of $M(0)$ is a measure of the magnitude of the QCD scale anomaly in $n=1$-point Schwinger functions \cite{Roberts:2016vyn};
and experiments on $Q^2\in [0,12]\,$GeV$^2$ at the upgraded JLab facility will be sensitive to the momentum dependence of $M(p)$ within a domain that is here indicated approximately by the shaded region.}
\end{figure}

Dynamical chiral symmetry breaking (DCSB) is a critical emergent phenome\-non in QCD, expressed in hadron wave functions, not in vacuum condensates \cite{Brodsky:2009zd,Brodsky:2010xf,Chang:2011mu,Brodsky:2012ku,Roberts:2015lja}; and contemporary theory argues that DCSB is responsible for more than 98\% of the visible mass in the Universe.  Given that classical massless-QCD is a conformally invariant theory, this means that DCSB is fundamentally connected with the origin of \emph{mass from nothing}.  The effect is evident in the dressed-quark propagator:
\begin{equation}
\label{Spgen}
S(p) = 1/[i \gamma\cdot p A(p^2) + B(p^2)] = Z(p^2)/[i\gamma\cdot p + M(p^2)]\,,
\end{equation}
where $M(p^2)$ is the dressed-quark mass-function, the behaviour of which is depicted and explained in Fig.\,\ref{gluoncloud}.  It is important to insist on the term ``dynamical,'' as distinct from spontaneous, because nothing is added to QCD in order to effect this remarkable outcome and there is no simple change of variables in the QCD action that will make it apparent.  Instead, through the act of quantising the classical chromodynamics of massless gluons and quarks, a large mass-scale is generated.

DCSB is very clearly revealed in properties of the pion, whose structure is described by a Bethe-Salpeter amplitude:
\begin{align}
\nonumber
\Gamma_{\pi}(k;P) & = \gamma_5 \left[
i E_{\pi}(k;P) + \gamma\cdot P F_{\pi}(k;P)  \right.\\
& \quad \left. +\, \gamma\cdot k \, G_{\pi}(k;P) + \sigma_{\mu\nu} k_\mu P_\nu H_{\pi}(k;P) \right],
\label{genGpi}
\end{align}
where $k$ is the relative momentum between the  valence-quark and -antiquark constituents (defined here such that the scalar functions in Eq.\,\eqref{genGpi} are even under $k\cdot P \to - k\cdot P$) and $P$ is their total momentum.  $\Gamma_{\pi}(k;P)$ is simply related to an object that would be the pion's Schr\"odinger wave function if a nonrelativistic limit were appropriate.  In QCD, if, and only if, chiral symmetry is dynamically broken, then one has in the chiral limit \cite{Binosi:2016wcx}:
\begin{equation}
\label{gtrE}
f_\pi^0 E_\pi(k;0) = B(k^2)\,,
\end{equation}
where $f_\pi^0$ is the chiral-limit value of the pion's leptonic decay constant.
This identity is remarkable. 
It is true in any covariant gauge, independent of the renormalisation scheme; and it means that the two-body problem is solved, nearly completely, once the solution to the one body problem is known.  Eq.\,\eqref{gtrE} is a quark-level Goldberger-Treiman relation.  It is also the most basic expression of Goldstone's theorem in QCD, \emph{viz}.\\[-3ex]
\hspace*{3em}\parbox[t]{0.9\textwidth}{\flushleft \emph{Goldstone's theorem is fundamentally an expression of equivalence between the one-body problem and the two-body problem in QCD's colour-singlet pseudoscalar channel}.}
\smallskip

\hspace*{-\parindent}Consequently, pion properties are an almost direct measure of the dressed-quark mass function depicted in Fig.\,\ref{gluoncloud}.  Thus, enigmatically, the qualities of the nearly-massless pion are the cleanest expression of the mechanism that is responsible for nearly all visible mass in the Universe.

It is now possible to resolve the dichotomy expressed by Eqs.\,\eqref{anomalyproton} and \eqref{anomalypion}.  These statements hold with equal force on a sizeable neighbourhood of the chiral limit because hadron masses are continuous functions of the current-quark masses.  So consider that for any meson, $M$,
\begin{equation}
\label{sigmaterm}
s_M(0) = \langle M(q) | m \bar \psi \psi | M(q) \rangle = m \frac{\partial m_M^2}{\partial m} \,,
\end{equation}
\emph{viz}.\ the scalar form factor at zero momentum transfer measures the response of the meson's mass-\emph{squared} to a change in current-quark mass and it is simply a \emph{convention} to define $\sigma_M = s_M(0)/[2 m_M]$.  Notably, the pion (and any other Nambu-Goldstone mode) possesses the peculiar property that
\begin{equation}
s_\pi(0) \stackrel{m\simeq 0}{=} m \frac{\partial m_\pi^2 }{\partial m} = 1 \times m_\pi^2 ,
\end{equation}
which is the statement that in the neighbourhood of the chiral limit, 100\% of the pion mass-squared owes to the current-mass in ${\mathpzc L}_{\rm QCD}$.  One should compare this result with that for the pion's spin-flip partner, \emph{i.e}.\ the $\rho$-meson \cite{Flambaum:2005kc}: $s_\rho(0)  \approx 0.06 \, m_\rho^2$, indicating that just 6\% of the $\rho$-meson's mass-squared is generated by the current-mass term in ${\mathpzc L}_{QCD}$.
Crucially, in neither of these cases, nor any others, are the current-quark masses alone responsible for the mass contributions described here.  Instead, they are obtained from the product of the small Higgs-generated current-quark masses with a DCSB-induced sQCD enhancement factor, \emph{viz}.\ the $Q^2=0$ value of the associated hadron's scalar form factor, its in-hadron condensate \cite{Chang:2011mu}.

The key to understanding Eq.\,\eqref{anomalypion} is Eq.\,\eqref{gtrE} and three partner Gold\-berger-Treiman-like relations, which are exact in chiral QCD.  Using these identities when working with those equations in quantum field theory that are necessary to describe a pseudoscalar bound state, one can construct an algebraic proof \cite{Binosi:2016rxz} that at any and each order in a symmetry-preserving analysis there is a precise cancellation between the mass-generating effect of dressing the valence-quarks which constitute the system and the attraction generated by the interactions between them.  This cancellation guarantees that the seed two-valence-parton system, which began massless, becomes a complex system, with a nontrivial bound-state wave function attached to a pole in the scattering matrix that is located at $P^2=0$.  Namely,  Eq.\,\eqref{anomalypion} is obtained through cancellations between one-body dressing and two-body binding effects:
\begin{align}
M^{\rm dressed}_{\rm quark} + M^{\rm dressed}_{\rm antiquark} + U^{\rm dressed}_{\rm quark-antiquark\;interaction} \stackrel{\rm chiral\;limit}{\equiv} 0\,,
\end{align}
with the sum being precisely zero if, and only if, chiral symmetry is dynamically broken in the Nambu pattern.

\section{Observing Mass in Baryons}
The empirical signals for confinement and mass generation are ubiquitous, but their manifestations do not typically appear the same in different systems.  Expressed otherwise, it is essential to study a diverse array of systems and/or observables -- spectrum of hadrons, hadron elastic and transition form factors, distribution amplitudes and functions, in all their guises, and the origin of the nucleon-nucleon interaction and emergence of nuclei -- because each one can potentially expose different aspects of the underlying mechanisms.  Then, development of a unified explanation of all observables will enable a complete picture of QCD to be drawn.

\subsection{Diquark Correlations}
This workshop focused on the spectrum and interactions of nucleon resonances; and highlighted just how crucial it has become to address the three valence-quark bound-state problem in QCD with the same level of sophistication that is now available for mesons \cite{Chang:2011vu,Horn:2016rip}.
In this connection, DCSB, evident in the momentum-dependence of the dressed-quark mass function -- Fig.\,\ref{gluoncloud}; is just as important to baryons as it is to mesons.
Indeed, one important consequence of DCSB is that any interaction capable of creating pseudo--Nambu-Goldstone modes as bound-states of a light dressed-quark and -antiquark, and reproducing the measured value of their leptonic decay constants, will necessarily also generate strong colour-antitriplet correlations between any two dressed-quarks contained within a nucleon. 
This assertion is based upon an accumulated body of evidence, gathered in two decades of studying two- and three-body bound-state problems in hadron physics \cite{Segovia:2015ufa}.  No realistic counter examples are known. 

The properties of such diquark correlations have been charted.  As color-carrying correlations, diquarks are confined \cite{Bender:1996bb,Bhagwat:2004hn}.  Additionally, owing to properties of charge-conjugation, a diquark with spin-parity $J^P$ may be viewed as a partner to the analogous $J^{-P}$ meson \cite{Cahill:1987qr}.  It follows that the strongest diquark correlations are: scalar isospin-zero, $[ud]_{0^+}$; and pseudovector, isospin-one, $\{uu\}_{1^+}$, $\{ud\}_{1^+}$, $\{dd\}_{1^+}$.  Moreover, whilst no pole-mass exists, the following mass-scales, which express the strength and range of the correlation, may be associated with these diquarks \cite{Cahill:1987qr,Maris:2002yu,Eichmann:2016yit,Eichmann:2016hgl,Lu:2017cln,Chen:2017pse} (in GeV):
\begin{equation}
m_{[ud]_{0^+}} \approx 0.7-0.8\,,\;
m_{\{uu\}_{1^+}}  \approx 0.9-1.1  \,,
\end{equation}
with $m_{\{dd\}_{1^+}}=m_{\{ud\}_{1^+}} = m_{\{uu\}_{1^+}}$ in the isospin symmetric limit.  The ground-state nucleon necessarily contains both scalar-isoscalar and pseudo\-vec\-tor-isovector correlations: neither can be ignored and their presence has many observable consequences \cite{Roberts:2013mja,Segovia:2013uga,Mezrag:2017znp}.  (Odd-parity baryons also contain pseudoscalar and vector diquarks, which may also play a role in even-parity excited states of the nucleon -- see, \emph{e.g}.\ Refs.\,\cite{Eichmann:2016yit,Eichmann:2016hgl,Lu:2017cln,Chen:2017pse} and the contributions of Bashir and Eichmann to these proceedings.)

Realistic diquark correlations are also soft and interacting.  All carry charge, scatter electrons, and possess an electromagnetic size which is similar to that of the analogous mesonic system, \emph{e.g}.\ \cite{Maris:2004bp,Eichmann:2008ef,Roberts:2011wy}:
\begin{equation}
\label{qqradii}
r_{[ud]_{0^+}} \gtrsim r_\pi, \quad r_{\{uu\}_{1^+}} \gtrsim r_\rho,
\end{equation}
with $r_{\{uu\}_{1^+}} > r_{[ud]_{0^+}}$.  As in the meson sector, these scales are set by that associated with DCSB.

\begin{figure}[t]
\centerline{%
\includegraphics[clip,width=0.7\textwidth]{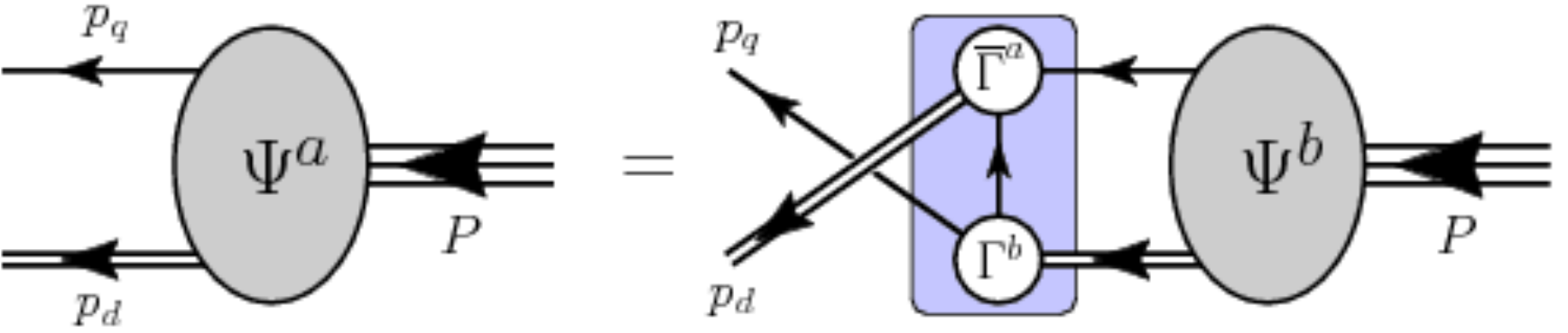}}
\caption{\label{figFaddeev}
Poincar\'e covariant Faddeev equation: a homogeneous linear integral equation for the matrix-valued function $\Psi$, being the Faddeev amplitude for a baryon of total momentum $P= p_q + p_d$, which expresses the relative momentum correlation between the dressed-quarks and -diquarks within the baryon.  The shaded rectangle demarcates the kernel of the Faddeev equation: \emph{single line}, dressed-quark propagator; $\Gamma$,  diquark correlation amplitude; and \emph{double line}, diquark propagator.}
\end{figure}

It is important to emphasise that these fully dynamical diquark correlations are vastly different from the static, pointlike ``diquarks'' which featured in early attempts \cite{Lichtenberg:1967zz,Lichtenberg:1968zz} to understand the baryon spectrum and explain the so-called missing resonance problem, \emph{viz}.\ the fact that quark models predict many more baryons states than were observed in the previous millennium \cite{Burkert:2004sk}.   As we have stated, modern diquarks are soft (not pointlike).  They also enforce certain distinct interaction patterns for the singly- and doubly-represented valence-quarks within the proton, as reviewed elsewhere \cite{Roberts:2015lja,Roberts:2013mja,Segovia:2014aza,Segovia:2016zyc}.  On the other hand, the number of states in the spectrum of baryons obtained from the Faddeev equation \cite{Eichmann:2016hgl,Lu:2017cln} is similar to that found in the three-constituent quark model.  (Notably, modern data and recent analyses have already reduced the number of missing resonances \cite{Ripani:2002ss,Burkert:2012ee,Kamano:2013iva,Crede:2013sze,Mokeev:2015moa,Anisovich:2017pmi}.)

The existence of these tight correlations between two dressed quarks is the key to transforming the three valence-quark scattering problem into the simpler Faddeev equation problem illustrated in Fig.\,\ref{figFaddeev}, without loss of dynamical information \cite{Eichmann:2009qa}.   The three gluon vertex, a signature feature of QCD's non-Abelian character, is not explicitly part of the bound-state kernel in this picture.  Instead, one capitalises on the fact that phase-space factors materially enhance two-body interactions over $n\geq 3$-body interactions and exploits the dominant role played by diquark correlations in the two-body subsystems.  Then, whilst an explicit three-body term might affect fine details of baryon structure, the dominant effect of non-Abelian multi-gluon vertices is expressed in the formation of diquark correlations.  Consequently, the active kernel here describes binding within the baryon through diquark breakup and reformation, which is mediated by exchange of a dressed-quark; and such a baryon is a compound system whose properties and interactions are largely determined by the quark$+$diquark structure evident in Fig.\,\ref{figFaddeev}.

\begin{figure*}[!t]
\hspace*{-2.5em}\begin{tabular}{ccc}
\parbox[c]{0.31\linewidth}{\includegraphics[clip,width=\linewidth]{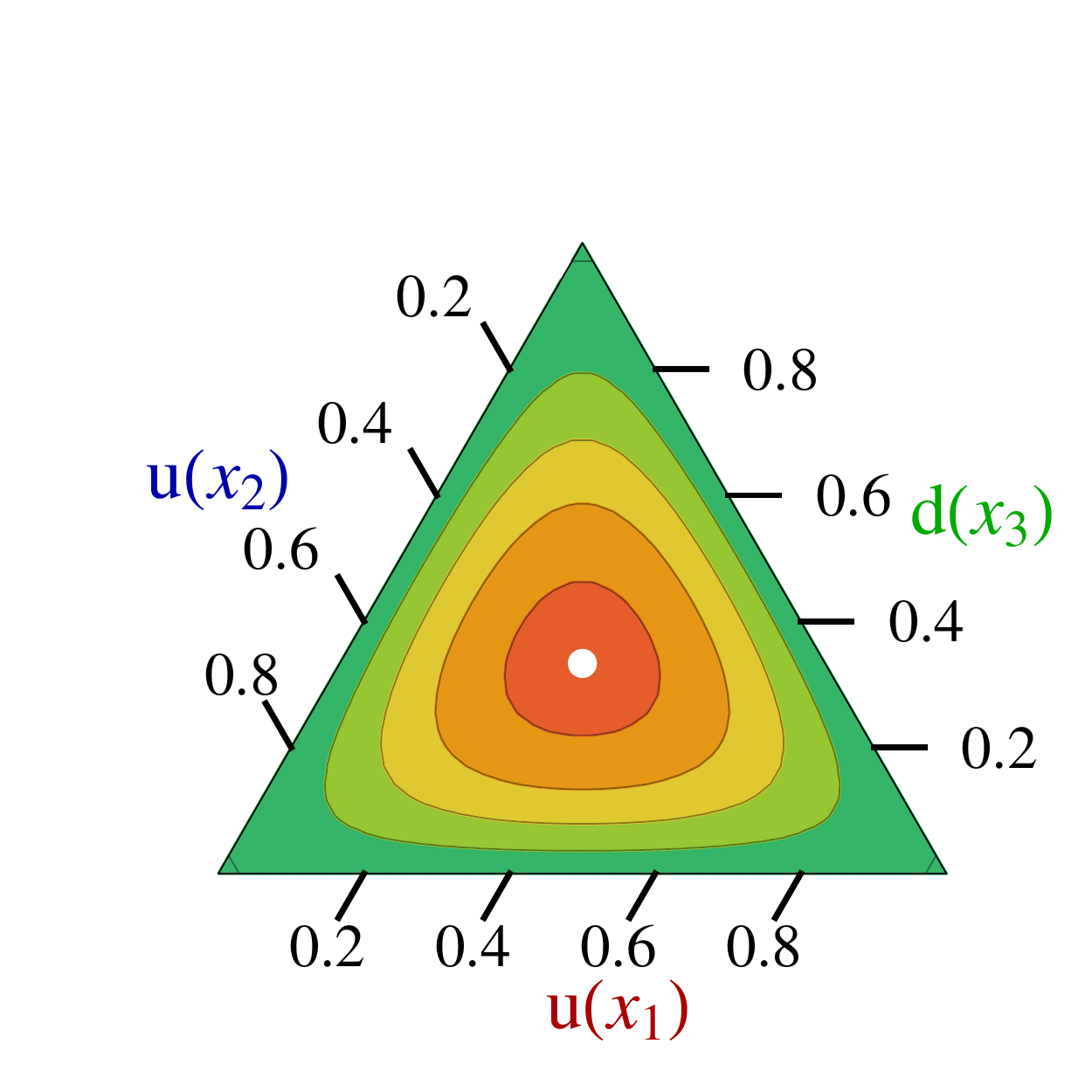}} &
\parbox[c]{0.31\linewidth}{\includegraphics[clip,width=\linewidth]{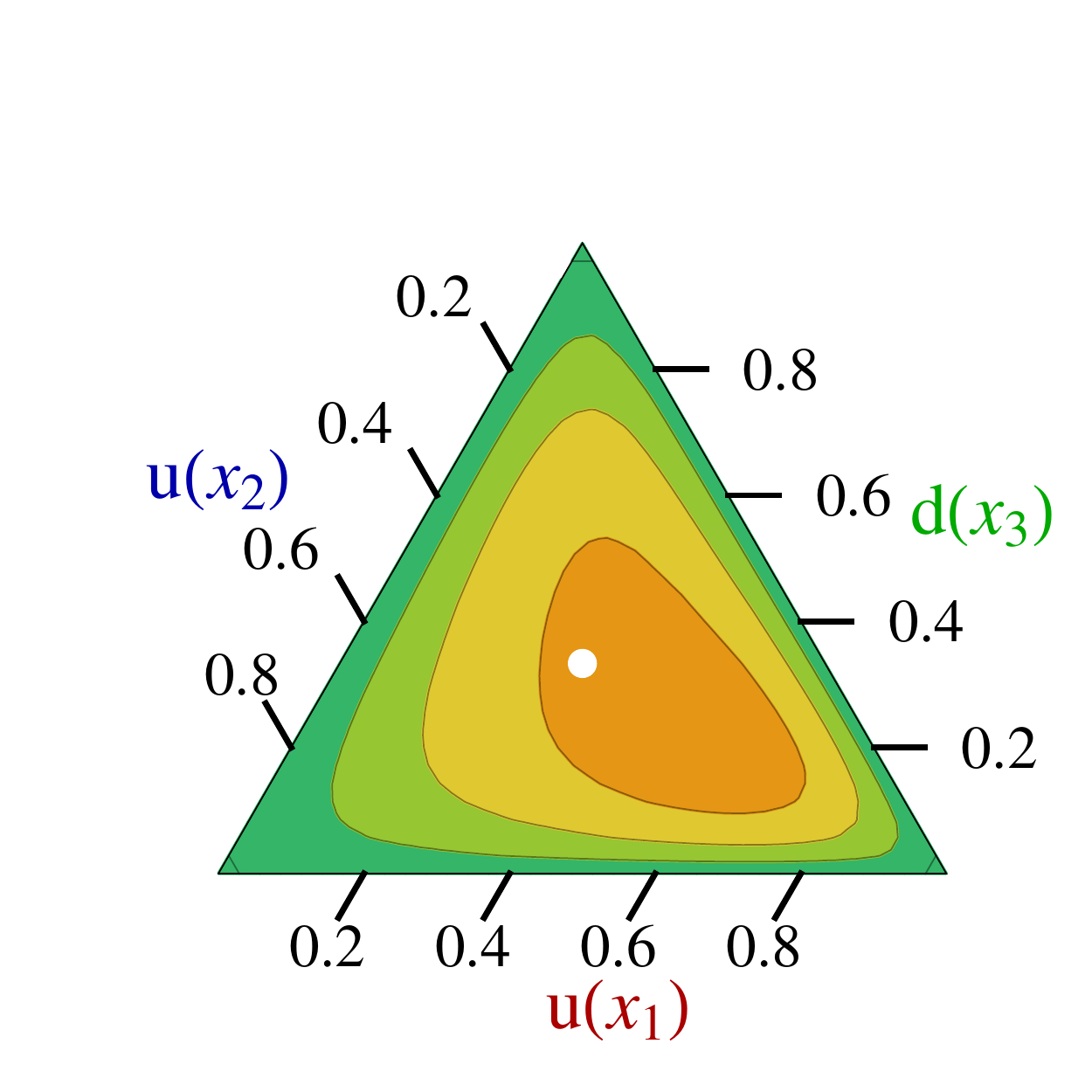}} &
\parbox[c]{0.37\linewidth}{\includegraphics[clip,width=\linewidth]{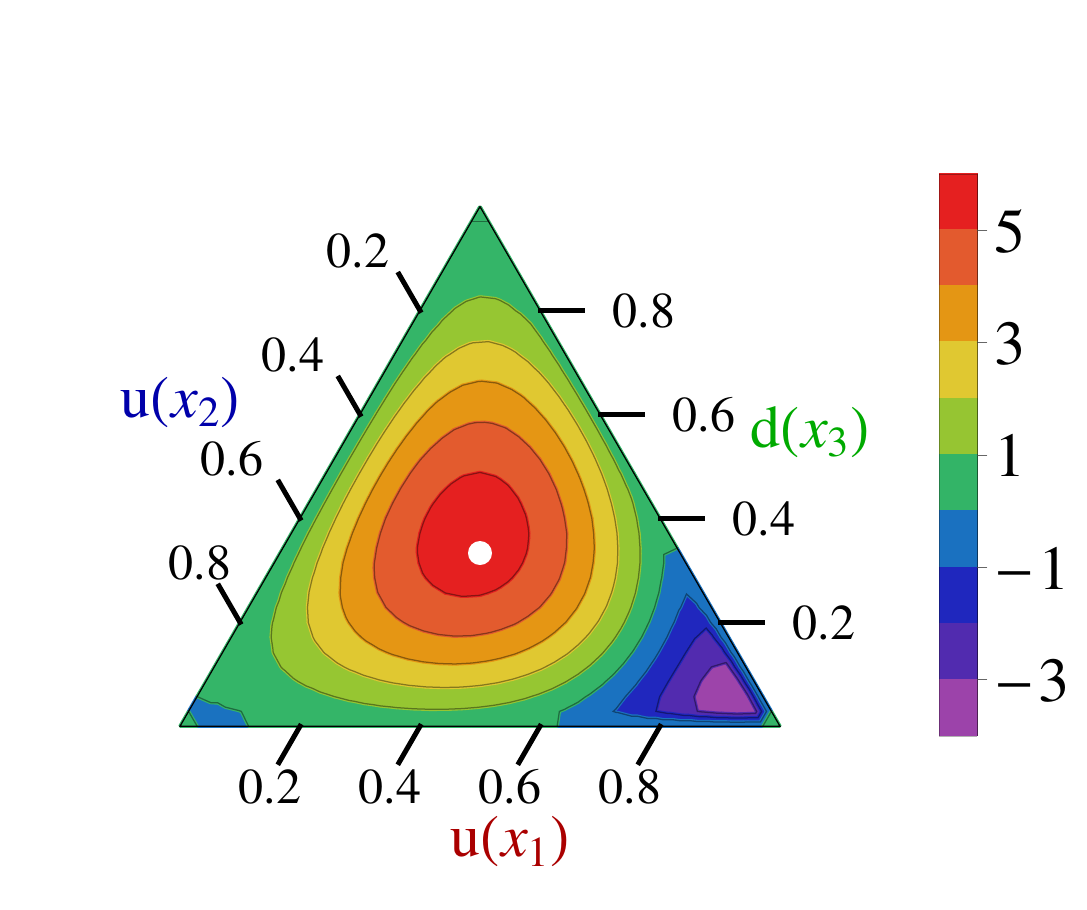}}
\end{tabular}\vspace*{-2ex}
\caption{\label{PlotPDAs} Barycentric plots:
\emph{left panel} -- conformal limit baryon PDA, $\varphi_N^{\rm cl}([x])=120 x_1 x_2 x_3$; \emph{middle panel} -- computed proton PDA evolved to $\zeta=2\,$GeV, which peaks at $([x])=(0.55,0.23,0.22)$; and \emph{right panel} -- Roper resonance PDA at $\zeta=2\,$GeV.  The white circle in each panel serves only to mark the centre of mass for the conformal PDA, whose peak lies at $([x])=(1/3,1/3,1/3)$.
}
\end{figure*}

Once the proton's Poincar\'e-covariant wave function, $\chi$, is known, \emph{e.g}.\ after solving the Faddeev equation in Fig.\,\ref{figFaddeev}, one can compute its dressed-quark leading-twist parton distribution amplitude (PDA). This is described in Ref.\,\cite{Mezrag:2017znp}, which used an algebraic \emph{Ansatz} for $\chi$, expressing the structural characteristics described above, to obtain the result depicted in Fig.\,\ref{PlotPDAs}.  For practical use, the numerical result is efficiently interpolated using ($w_{00}=1$)
{\allowdisplaybreaks
\begin{eqnarray}
\nonumber
\varphi([x]) & = & {\mathpzc n}_{\varphi} \, x_1^{\alpha_-} (x_2 x_3)^{\beta_-}
\sum_{j=0}^2\sum_{i=0}^j w_{ij} \, P_{j-i}^{2[i+\beta];\alpha_-}(2x_1-1) \\
&& \times  (x_2+x_3)^i C_i^\beta([x_3-x_2]/[x_2+x_3])\,, \label{EqInterpolation}
\end{eqnarray}
}%
\hspace*{-0.0\parindent}where $(\alpha,\beta)_- = (\alpha,\beta)-1/2$;
$P$ is a Jacobi function, $C$ a Gegenbauer polynomial; the interpolation parameters are listed in Table\,\ref{interpolation}A; and ${\mathpzc n}_{\varphi}$ ensures
\begin{align}
 \int _0^1 dx_1 dx_2 dx_3 \delta(1-\sum_i x_i)\varphi([x]) & =: \int [dx] \varphi([x])=  1\,.
\end{align}

\begin{table}[!t]
\caption{\emph{A} --  Eq.\,\eqref{EqInterpolation} interpolation parameters for the proton and Roper PDAs in Fig.\,\ref{PlotPDAs}.  
%
\emph{B} -- Computed values of the first four moments of the PDAs.  The error on $f_N$, a dynamically-determined quantity which measures the proton's ``wave function at the origin'', reflects a nucleon scalar diquark content of $65\pm 5$\%; and values in rows marked with ``$\not\supset \mbox{av}$'' were obtained assuming the baryon is constituted solely from a scalar diquark.
(All results listed at a renormalisation scale $\zeta=2\,$GeV.)
\label{interpolation}
}
\begin{tabular*}
{\hsize}
{
l|@{\extracolsep{0ptplus1fil}}
c|@{\extracolsep{0ptplus1fil}}
c|@{\extracolsep{0ptplus1fil}}
c|@{\extracolsep{0ptplus1fil}}
c|@{\extracolsep{0ptplus1fil}}
c|@{\extracolsep{0ptplus1fil}}
c|@{\extracolsep{0ptplus1fil}}
c|@{\extracolsep{0ptplus1fil}}
c@{\extracolsep{0ptplus1fil}}}\hline
A & ${\mathpzc n}_{\hat\varphi}$ & $\alpha$ & $\beta$ & $w_{01}$ & $w_{11}$ & $w_{02}$ & $w_{12}$ & $w_{22}$ \\\hline
$p$ & 65.8 & 1.47 & 1.28 & $\phantom{-}0.096$ & $0.094$ & $\phantom{-}0.15$ & $-0.053$ & $\phantom{-}0.11$ \\
$R$ & 14.4 & 1.42 & 0.78 & $-0.93\phantom{6}$ & $0.22\phantom{0}$ & $-0.21$ & $-0.057$ & $-1.24$ \\\hline
\end{tabular*}
\smallskip

\begin{tabular*}
{\hsize}
{
l|@{\extracolsep{0ptplus1fil}}
c|@{\extracolsep{0ptplus1fil}}
c|@{\extracolsep{0ptplus1fil}}
c|@{\extracolsep{0ptplus1fil}}
c@{\extracolsep{0ptplus1fil}}}\hline
B & $10^3 f_N/\mbox{GeV}^2$ & $\langle x_1\rangle_u$ & $\langle x_2\rangle_u$ & $\langle x_3\rangle_d$ \\\hline
conformal PDA & & $0.333\phantom{(99)}$ & $0.333\phantom{(9)}$ & $0.333\phantom{(9)}$ \\\hline
lQCD \mbox{\cite{Braun:2014wpa}} & $2.84(33)$ & $0.372(7)\phantom{9}$ & 0.314(3) & 0.314(7) \\
lQCD \mbox{\cite{Bali:2015ykx}} & $3.60(6)\phantom{9}$ &  $0.358(6)\phantom{9}$ & $0.319(4)$ & 0.323(6) \\\hline
herein proton & $3.78(14)$ & $0.379(4)\phantom{9}$ & 0.302(1) & 0.319(3) \\
herein proton $\not\supset \mbox{av}$ & $2.97\phantom{(17)}$ & $0.412\phantom{(17)}$ & $0.295\phantom{(7)}$ & $0.293\phantom{(7)}$ \\\hline\hline
herein Roper  & $5.17(32)$ & $0.245(13)$ & $0.363(6)$ & $0.392(6)$ \\
herein Roper $\not\supset \mbox{av}$ & $2.63\phantom{(14)}$ & $0.010\phantom{(19)}$ & $0.490\phantom{(9)}$ & $0.500\phantom{(9)}$ \\\hline
\end{tabular*}
\end{table}

Table\,\ref{interpolation}B lists the four lowest-order moments of the proton PDA.  They reveal valuable insights, \emph{e.g}.\ when the proton is drawn as solely a quark+sca\-lar-diquark correlation, $\langle x_2 \rangle_u=\langle x_3 \rangle_d$, because these are the two participants of the scalar quark+quark correlation; and the system is very skewed, with the PDA's peak being shifted markedly in favour of $\langle x_1 \rangle_u > \langle x_2 \rangle_u$.  This outcome conflicts with lQCD results \cite{Braun:2014wpa,Bali:2015ykx}.  On the other hand, as described above, realistic Faddeev equation calculations indicate that pseudovector diquark correlations are an essential part of the proton's wave function.  Naturally, when these $\{uu\}$ and $\{ud\}$ correlations are included, momentum is shared more evenly, shifting from the bystander $u(x_1)$ quark into $u(x_2)$, $d(x_3)$.  Adding these correlations with the known weighting, the PDA's peak moves back toward the centre, locating at $([x])=(0.55,0.23,0.22)$; and the computed values of the first moments align with those obtained using lQCD.  This confluence delivers a significantly more complete understanding of the lQCD simulations, which are thereby seen to validate a picture of the proton as a bound-state with both strong scalar \emph{and} pseudovector diquark correlations, in which the scalar diquarks are responsible for $\approx 60$\% of the Faddeev amplitude's canonical normalisation.

Importantly, as found with ground-state $S$-wave mesons \cite{Horn:2016rip,Chang:2013pq,Shi:2015esa,Braun:2015axa,Gao:2016jka,Zhang:2017bzy,Chen:2017gck}, the leading-twist PDA of the ground-state nucleon is both broa\-der than $\varphi_N^{\rm cl}([x])$ and decreases monotonically away from its maximum in all directions, \emph{i.e}.\ the PDAs of these ground-state $S$-wave systems possess endpoint enhancements, but neither humps nor bumps.
Models which produce such structures were previously considered reasonable \cite{Chernyak:1983ej}.  However, it is now evident that pointwise behaviour of this type is inconsistent with QCD.  The models may nevertheless be viewed as possessing a qualitative reality, insofar as they represent a means by which endpoint enhancements can be introduced into hadron PDAs if one restricts oneself to the basis characterising QCD's conformal limit.

The Faddeev equation framework has also been employed to determine the properties of the dressed-quark core of the Roper resonance, the nucleon's first radial excitation \cite{Chen:2017pse,Segovia:2015hra}; and, \emph{e.g}.\ the scalar functions in this system's Faddeev amplitude possess a zero at quark-diquark relative momentum $\surd \ell^2 \approx 0.4\,$GeV$\approx 1/[0.5\,{\rm fm}]$.  Based on those results, Ref.\,\cite{Mezrag:2017znp} delivered the associated leading-twist PDA for this system, depicted in the rightmost panel of Fig.\,\ref{PlotPDAs}, which is efficiently interpolated using Eq.\,\eqref{EqInterpolation} with the parameters in Table\,\ref{interpolation}A; and whose first four moments are listed in  Table\,\ref{interpolation}B.  This prediction reveals some curious features, \emph{e.g}.: the excitation's PDA is not positive definite and there is a prominent locus of zeros in the lower-right corner of the barycentric plot, both of which echo features of the wave function for the first radial excitation of a quantum mechanical system and have also been seen in the leading-twist PDAs of radially excited mesons \cite{Li:2016dzv,Li:2016mah}; and the impact of pseudovector correlations within this excitation is opposite to that in the ground-state, \emph{viz}.\ they shift momentum into $u(x_1)$ from $u(x_2)$, $d(x_3)$.

The veracity of these PDA predictions can be tested in future studies.  For instance, they will enable the first realistic assessments of the scale at which exclusive experiments involving baryons may properly be compared with predictions based on perturbative-QCD hard scattering formulae and thereby assist contemporary and planned facilities to refine and reach their full potential \cite{Accardi:2012qut,Dudek:2012vr,Burkert:2016dxc}.  The value of such estimates has recently been demonstrated in studies of mesons \cite{Horn:2016rip,Chang:2013nia,Gao:2017mmp}.

\subsection{Transition Form Factors}
The prediction and measurement of ground-state elastic form factors is essential to increasing our understanding of sQCD; and there have already been surprising discoveries, \emph{e.g}.\ Refs.\,\cite{Jones:1999rz,Gayou:2001qd,Punjabi:2005wq,Puckett:2010ac,Puckett:2011xg}.  Alone, however, it is insufficient to explore and expose the infrared behaviour of the strong interaction.  (The hydrogen ground-state did not deliver QED!)  As this workshop highlighted, there are numerous nucleon$\to$resonance transition form factors; and the challenges of mapping and explaining their $Q^2$-dependence provide a vast array of novel ways to delve into QCD, including: charting the transition from the perturbative to the strongly interacting domain (already highlighted in Fig.\,\ref{gluoncloud}); and the environment sensitivity of correlations.

\begin{figure}[t]
\centerline{\includegraphics[width=1.0\textwidth]{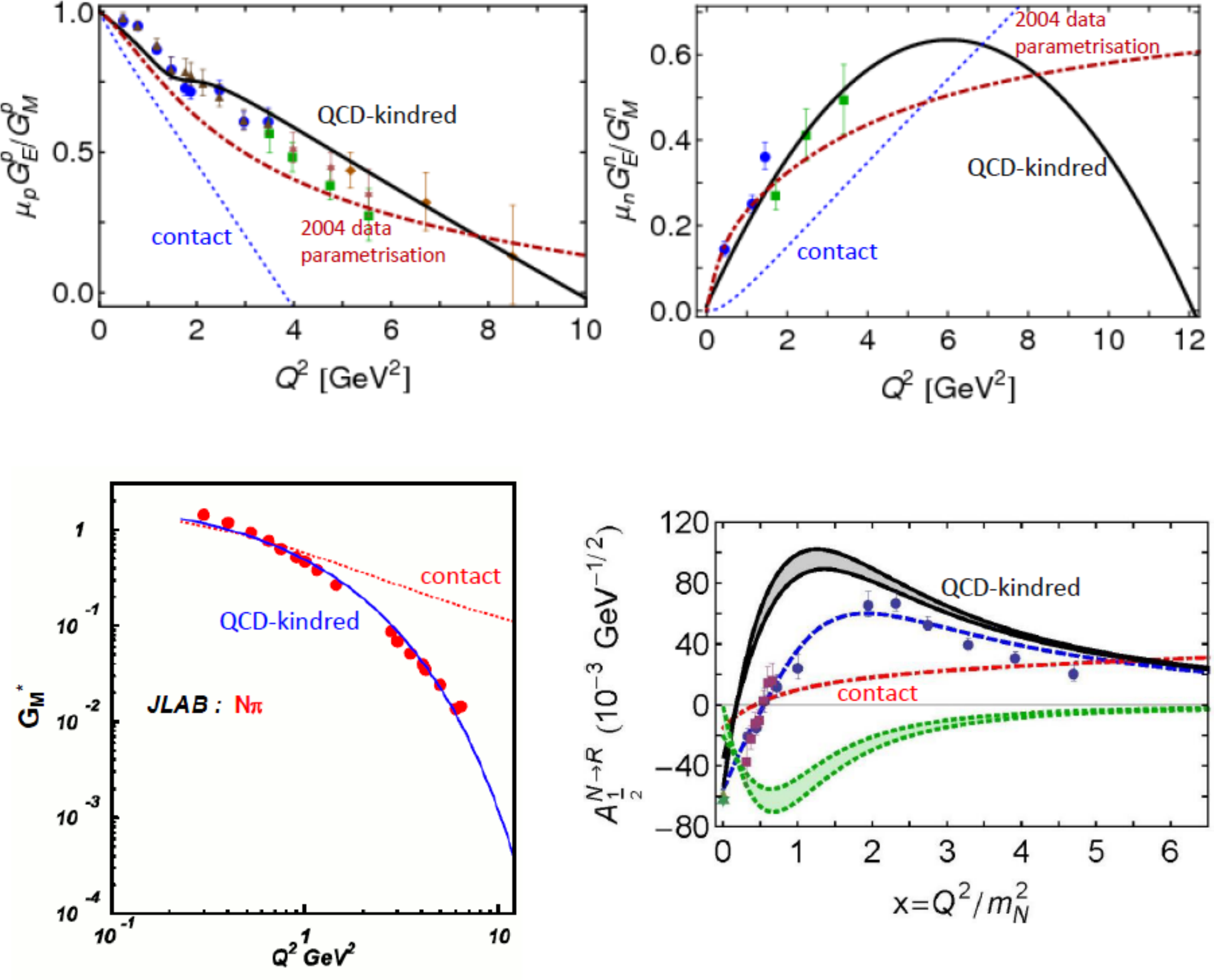}}
\caption{\label{contactkindred}
For a diverse range of baryon properties, these four panels compare results obtained using an internally-consistent treatment of a vector$\,\otimes\,$vector contact interaction with the predictions of a widely-used QCD-kindred Faddeev-equation framework \cite{Segovia:2014aza}.
The dot-dashed (red) curve in the upper two panels depicts a parametrisation of pre-2004 data \cite{Kelly:2004hm}; the data in the lower-left panel is from Ref.\,\cite{Aznauryan:2009mx}; and the dashed (blue) curve in the lower-right panel is a fit to the experimental data (circles (blue) \cite{Aznauryan:2009mx}; squares (purple) \cite{Mokeev:2012vsa,Mokeev:2015lda}; and star (green) \cite{Olive:2016xmw}).
}
\end{figure}

In relativistic quantum field theory, once the quark-quark interaction is specified, then all elements in a calculation of hadron properties are fully determined, \emph{e.g}.\ one-body propagators, two-body correlation amplitudes, bound-state wave functions, \ldots, and hence every hadron observable.  There is no freedom to ``fiddle'': all elements in a calculation are correlated and one piece cannot be changed without each of the others responding, too, in some (mathematically determined) manner.
Therefore, if one treats two different interactions on an equal footing, then it is possible to determine whether and how experiment can distinguish between them.  

This last point is illustrated in Fig.\,\ref{contactkindred}, which, for an array of observables, displays a comparison between predictions made by the QCD-based framework in Ref.\,\cite{Segovia:2014aza} and results obtained using a confining, symmetry-preserving treatment of a vector$\,\otimes\,$vector contact interaction \cite{Segovia:2013uga,Wilson:2011aa,Segovia:2013rca,Xu:2015kta}.  Any such treatment of a contact-interaction produces hard hadron form factors, curtails some quark orbital angular momentum correlations within a baryon, and suppresses two-loop diagrams in the baryon elastic and transition electromagnetic currents.\footnote{Typical formulations of Nambu-Jona-Lasinio models deliver an inconsistent treatment of the contact interaction, \emph{e.g}.\ they suppress dynamically-required components of a pseudoscalar meson's bound-state amplitude which dominate the $Q^2>M(0)^2$ behaviour of form factors, thereby artificially obtaining soft form factors \cite{GutierrezGuerrero:2010md, Roberts:2010rn, Roberts:2011wy, Wilson:2011aa, Chen:2012txa, Segovia:2013uga, Xu:2015kta, Bedolla:2016yxq}.}
These defects are rectified in the QCD-based approach so that comparisons, such as those illustrated in Fig.\,\ref{contactkindred}, expose those quantities which are most sensitive to the momentum dependence of elementary quantities in QCD and where to look for that sensitivity.

It is apparent from Fig.\,\ref{contactkindred} that much existing data supports the notion that the strong-interaction is described by a $1/k^2$ vector$\,\otimes\,$vector interaction, but the effects and $Q^2$-domains change with the target.  The high $Q^2$ available at JLab\,12 will enable experiment to climb out of the meson-cloud region for a wide array of systems.\footnote{Meson-baryon final-state-interactions (MB\,FSIs) -- commonly called meson cloud effects -- have a significant impact on nucleon-to-resonance transitions, \emph{e.g}.\ on $Q^2\simeq 0$, they are responsible for approximately one-third of the magnetic $\gamma^\ast N \to \Delta$ transition form factor \cite{JuliaDiaz:2006xt} -- also consider the contributions of D\"oring, Kamano and Tiator to these proceedings.  Since strong interaction bound-states are soft, MB\,FSI effects diminish rapidly with increasing probe momenta.  For dominant amplitudes, they are typically negligible on $Q^2\gtrsim 2\,$GeV$^2$; and hence such hard probes provide access to a hadron's dressed-quark core.}
On this domain of quark-core dominance, the measurements will deliver clear signals for the momentum-dependence of the strong-interaction's effective coupling and masses because different power-laws for the interaction predict distinct scaling behaviour for the form factors: with any set of candidate interactions tuned to produce a few static hadron properties, their predictions for dynamical properties diverge significantly as, \emph{e.g}.\ the probe momentum increases.

\subsection{Roper Resonance}
An explanation of how and where the Roper resonance fits into the emerging spectrum of hadrons cannot rest on a description of its mass alone.  Instead, it must combine an understanding of the Roper's mass and width with a detailed account of its structure and how that structure is revealed in the momentum dependence of the transition form factors.  Furthermore, it must unify all this with a similarly complete picture of the proton from which the Roper resonance is produced.  This is a prodigious task.

After more than fifty years, however, a coherent picture connecting the Roper resonance with the nucleon's first radial excitation has become visible.  Completing this portrait only became possible following
acquisition and analysis of a vast amount of high-precision nucleon-resonance electroproduction data with single- and double-pion final states on a large kinematic domain of energy and momentum-transfer;
development of a sophisticated dynamical reaction theory capable of simultaneously describing all partial waves extracted from available, reliable data;
formulation and wide-ranging application of a Poincar\'e covariant approach to the continuum bound state problem in relativistic quantum field theory that expresses diverse local and global impacts of DCSB in QCD;
and the refinement of constituent quark models so that they, too, qualitatively incorporate these aspects of strong QCD.
In this picture:
\begin{itemize}
\setlength\itemsep{0em}
\item the Roper resonance is, at heart, the first radial excitation of the nucleon.
\item it consists of a well-defined dressed-quark core, which plays a role in determining the system's properties at all length-scales, but exerts a dominant influence on probes with $Q^2\gtrsim m_p^2$;
\item and this core is augmented by a meson cloud, which both reduces the Roper's core mass by approximately 20\%, thereby solving the mass problem that was such a puzzle in constituent-quark model treatments, and, at low-$Q^2$, contributes an amount to the electroproduction transition form factors that is comparable in magnitude with that of the dressed-quark core, but vanishes rapidly as $Q^2$ is increased beyond $m_p^2$.
\end{itemize}
This perspective is detailed elsewhere \cite{Burkert:2017djo} -- consider also the articles by Burkert, Carman and Mokeev in these proceedings.  Nevertheless, it is here worth describing an insightful analysis of an associated transition charge density.

The nucleon-$\Delta$ and nucleon-Roper transition form factors have been dissected in order to reveal the relative contributions from dressed-quarks and the various diquark correlations -- see Ref.\,\cite{Segovia:2016zyc} and Segovia's contribution to these proceedings.  This analysis reveals that $F_1^\ast$ is largely determined by a process in which the virtual photon scatters from the uncorrelated $u$-quark with a $[ud]_{0^+}$ diquark as a spectator, with lesser but non-negligible contributions from other processes.  In exhibiting these properties, $F_1^\ast$ shows qualitative similarities to the proton's Dirac form factor.

Such features of the transition can also be highlighted by studying the following transition charge density \cite{Tiator:2008kd}:
\begin{align}
\label{eqrhob}
\rho^{pR}(|\vec{b}|)
& := \int \frac{d^2 \vec{q}_\perp }{(2\pi)^2} \,{\rm e}^{i \vec{q}_\perp \cdot \vec{b}} F_1^\ast(|\vec{q}_\perp|^2)\,,
%
\end{align}
where $F_1^\ast$ is the proton-Roper Dirac transition form factor, interpreted in a frame defined by $Q=({q}_\perp=(q_1,q_2),Q_3=0,Q_4=0)$.  Plainly, $Q^2 = |\vec{q}_\perp|^2$.  Defined in this way, $\rho^{pR}(|\vec{b}|)$ is a light-front-transverse charge-density with a straightforward quantum mechanical interpretation \cite{Miller:2007uy}.  Notably,
\begin{equation}
\int d^2\vec{b}\,\rho^{pR}(|\vec{b}|)  = F_1^\ast(0) = 0\,.
\end{equation}

\begin{figure}[t]
\begin{center}
\begin{tabular}{lr}
\includegraphics[clip,width=0.47\linewidth]{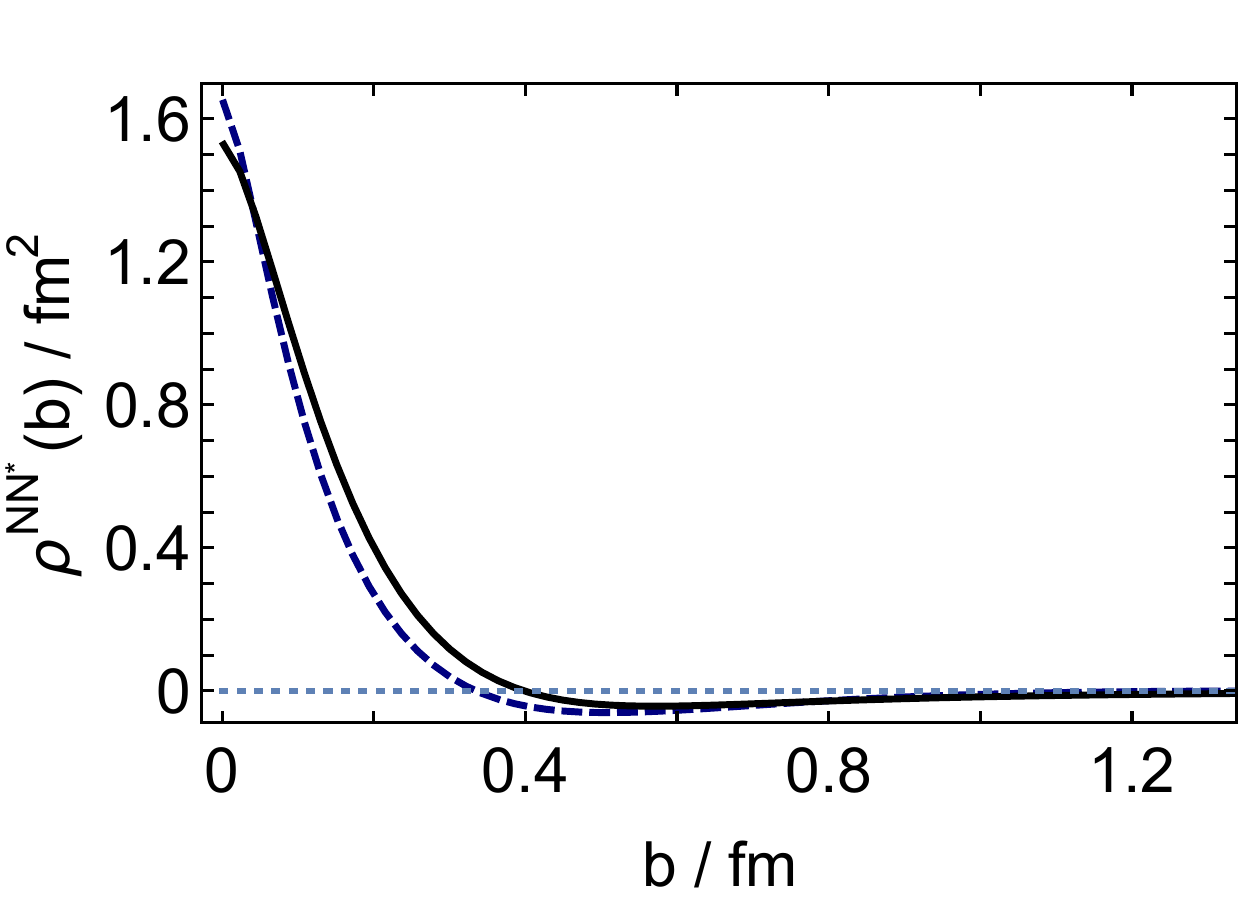} &
\includegraphics[width=0.46\linewidth]{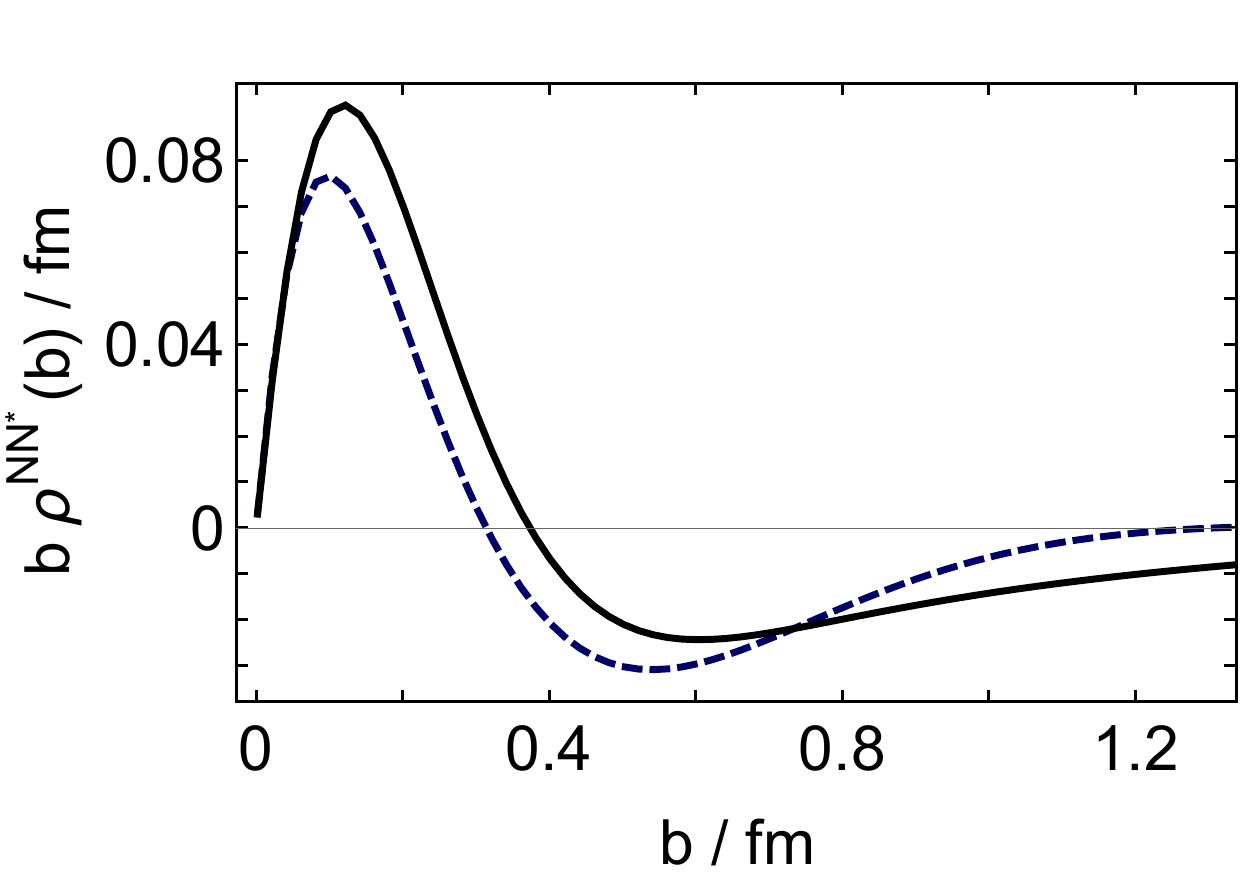}
\end{tabular}
\end{center}
%
%
\caption{\label{rhob}
$\rho^{pR}(|\vec{b}|)$ (left panel) and $|\vec{b}| \rho^{pR}(|\vec{b}|) $ (right) calculated using Eq.\,\eqref{eqrhob}: solid (black) curve -- dressed-quark core contribution, computed using the midpoint-result within the gray bands in the upper-left panel of Fig.\,3, Ref.\,\cite{Roberts:2016dnb}; and dashed (blue) curve -- empirical result, computed using the dashed (blue) curve therein.}
\end{figure}

Fig.\,\ref{rhob} depicts a comparison between the empirical result for $\rho^{pR}(|\vec{b}|)$ and the dressed-quark core component: the difference between these curves measures the impact of MB\,FSIs on the transition.
Within the domain displayed, both curves describe a dense positive center, which is explained by noting that the proton-Roper transition is dominated by the photon scattering from a positively-charged $u$-quark in the presence of a positively-charged $[ud]_{0^+}$ diquark spectator, as mentioned above.
Furthermore, both curves exhibit a zero at approximately 0.3\,-\,0.4\,fm, with that of the core lying at larger $|\vec{b}|$.  Thence, after each reaching a global minimum, the dressed-quark core contribution returns slowly to zero from below whereas the empirical result returns to pass through zero once more, although continuing to diminish in magnitude.

The long-range negative tail of the quark core contribution, evident in Fig.\,\ref{rhob}, reveals the increasing relevance of axial-vector diquark correlations at large distances because the $d\{uu\}_{1^+}$ component is twice as strong as $u\{ud\}_{1^+}$ in the proton and Roper, and photon interactions with uncorrelated quarks dominate the transition.
Moreover, consistent with their role in reducing the nucleon and Roper quark-core masses, one sees that MB\,FSIs introduce significant attraction, working to screen the long negative tail of the quark-core contribution and thereby compressing the transition domain in transverse space.  (The dominant long-range MB effect is $n\pi^+$, which generates a positive tail.) As measured by the rms transverse radius, the size of the empirical transition domain is just two-thirds of that associated with the dressed-quark core.

These fifty years of experience with the Roper resonance have delivered important lessons.  Namely, in attempting to predict and explain the QCD spectrum, one must: fully consider the impact of MB\,FSIs, and the couplings between channels and states that they generate; and look beyond merely locating the poles in the $S$-matrix, which themselves reveal little structural information, to also consider the $Q^2$-dependences of the residues, which serve as a penetrating scale-dependent probe of resonance composition.

Moreover, the Roper resonance is not unusual.  Indeed, in essence, the picture described here is also applicable to the $\Delta$-baryon; and an accumulating body of experiment and theory indicates that almost all baryon resonances can be viewed the same way, \emph{viz}.\ as systems possessing a three-body dressed-quark bound-state core that is supplemented by a meson cloud, whose importance varies from state to state and whose observable manifestations disappear rapidly as the resolving power of the probe is increased.  In this connection, it is important to highlight that CLAS12 at the newly upgraded JLab will be capable of determining the electrocouplings of most prominent nucleon resonances at unprecedented photon virtualities: $Q^2\in [6,12]\,$GeV$^2$ \cite{E12-09-003,E12-06-108A}.  Consequently, the associated experimental program will be a powerful means of validating the perspective described herein.

\section{Epilogue}
Nucleon resonance physics, experiment and theory, has made significant pro\-gress in the past two decades.
Regarding ``missing resonances'', gaps are being filled, and the growing number of states increasingly fits the pattern expected of a three-body system, with correlations.
In addition, novel insights concerning the momentum-dependence of QCD's running coupling and masses have been drawn from data on nucleon elastic and transition form factors.
Electroproduction data on a wide $Q^2$-domain, reaching high-$Q^2$, has been crucial to this progress, revealing much about the internal structure of resonances and paving the way to a solution of the Roper resonance puzzle.

During the next decade, CLAS\,12 will deliver resonance electroproduction data out to $Q^2 \approx 12\,$GeV$^2$ and thereby empirical information which can address a wide range of issues that are critical to our understanding of strong interactions, \emph{e.g}.: is there an environment sensitivity of DCSB; and are quark-quark correlations an essential element in the structure of all baryons?  Existing experiment-theory feedback suggests that there is no environment sensitivity for the nucleon, $\Delta$-baryon and Roper resonance: DCSB in these systems is expressed in ways that can readily be predicted once its manifestation is understood in the pion, and this includes the generation of diquark correlations with the same character in each of these baryons.
Resonances in other channels, however, probably contain additional diquark correlations, with different quantum numbers, and can potentially be influenced in new ways by MB\,FSIs.  Therefore, these channels, and higher excitations, open new windows on sQCD and its emergent phenomena whose vistas must be explored and mapped if the most difficult part of the Standard Model is finally to be solved.

\begin{acknowledgements}
I would like to thank
R. Gothe, 
Y. Ilieva, 
V. Mokeev, 
E. Santopinto, 
and S. Strauch 
for their efforts in organising the 11th International Workshop on the Physics of Excited Nucleons, 20 -- 23 August 2017, at the University of South Carolina, Columbia, SC,
and their kindness and support during the meeting.
This contribution is based on results obtained through collaborations with many people, to all of whom I am greatly indebted.
Work supported by:
the U.S.\ Department of Energy, Office of Science, Office of Nuclear Physics, under contract no.~DE-AC02-06CH11357.
\end{acknowledgements}


\end{document}